\documentclass[preprint,aps,pre,showpacs,amsmath,amsfonts,amssymb,floatfix,superscriptaddress]{revtex4}
\usepackage{graphicx}
\usepackage{dcolumn}
\usepackage{bm}
\usepackage{amsmath,amsfonts,amssymb,dsfont,color,bbm,tikz,bbold}
\usepackage{epstopdf}
\usepackage{dsfont}

\newcommand{\bra}[1]{\left\langle #1\right|}
\newcommand{\ket}[1]{\left| #1\right\rangle}

\begin{document}

\title{Aspects of Floquet Bands and Topological Phase Transitions in a Continuously Driven Superlattice}
\author{Longwen Zhou}
\affiliation{Department of Physics and Center for Computational Science and Engineering,
National University of Singapore, 117542, Singapore}
\author {Hailong Wang}
\affiliation{Department of Physics and Center for Computational Science and Engineering,
National University of Singapore, 117542, Singapore}
\author {Derek Y.~H. Ho}
\affiliation{Department of Physics and Center for Computational Science and Engineering,
National University of Singapore, 117542, Singapore}
\author{Jiangbin Gong} \email{phygj@nus.edu.sg}
\affiliation{Department of Physics and Center for Computational Science and Engineering,
National University of Singapore, 117542, Singapore}
\affiliation{NUS Graduate School for Integrative Sciences and Engineering, Singapore
117597, Singapore}
\date{\today}

\begin{abstract}


Recently the creation of novel topological states of matter by a periodic driving field has attracted great attention. To motivate further experimental and theoretical studies,  we investigate interesting aspects of Floquet bands and topological phase transitions in a continuously driven Harper model.  In such a continuously driven system with an odd number of Floquet bands, the bands are found to have nonzero Chern numbers in general and topological phase transitions take place as we tune various system parameters, such as the amplitude or the period of the driving field. The nontrivial Floquet band topology results in a quantized transport of Wannier states in the lattice space.
For certain parameter choices, very flat yet topologically nontrivial Floquet bands may also emerge, a feature that is potentially useful for the simulation of physics of strongly correlated systems.
Some cases with an even number of Floquet bands may also have intriguing Dirac cones in the spectrum. Under open boundary conditions, anomalous counter-propagating chiral edge modes and degenerate zero modes  are also found as the system parameters are tuned. These results should be of experimental interest because a continuously driven system is easier to realize than a periodically kicked system.


\end{abstract}

\maketitle

\section{Introduction}

Floquet topological states of matter are found on the edges of periodically driven quantum systems which have nontrivial bulk band topology. In recent years the study of such states has attracted increasing theoretical and experimental interests.
Much progress has been made in this attractive new field and several fascinating
examples of exotic Floquet topological states of matter (e.g., Floquet quantum Hall effects~\cite{FQHDahlhausPRB2011,FQHDerekPRL2012,FQHHailongPRE2013,FQHOzawaPRL2014,FQHZhaoErHPRL2014}, Floquet graphene~\cite{FGrapheAokiPRB2009,FGrapheAuerbachPRL2011,FGrapheDelplacePRB2013,FGrapheIadecolaPRB2014,FGrapheIadecolaPRL2013,FGrapheKitagawaPRB2011,FGrapheKogheePRA2012,FGrapheLopezArx2013,FGrapheMorellPRB2012,FGrapheTorresPRB2014,FGrapheZhaiArx2014}, Floquet topological insulators~\cite{FTICayssolPRL2012,FTICayssolPSSRRL2013,FTIChongPRB2014,FTIGalitskiPRB2013,FTIInouePRL2010,FTIIurovJPCM2013,FTILindnerNatPhys2011,FTILindnerPRB2013,FTILumerPRL2013,FTINakagawaPRA2014,FTIPodolskyPRB2013,FTIPodolskyPRL2013,FTIRefaelArx2014}, Floquet Majorana fermions~\cite{FMFDuttaPRB2013,FMFGongPRB2013,FMFJiangPRL2011,FMFKunduPRL2013,FMFLiArx2014,FMFLiuPRL2013,FMFLiuWWEPL2013,FMFReynosoJPCM2014,FMFReynosoPRB2013,FMFSatoArx2014,FMFThakurathiArx2013,FMFTrifPRL2012,FMFXieArx2014}) have been proposed theoretically or even realized experimentally~\cite{FloqExpGedikSci2013,FloqExpKitagawaNatCom2012,FloqExpPuentesPRL2014,FloqExpRechtsmanNat2013I,FloqExpRechtsmanPRL2013II}. Their potential applications reside in spintronics, quantum information processing and quantum computation.

The introduction of a periodic driving field may renormalize a particular part of a Hamiltonian, induce topologically nontrivial quasienergy winding, and alter the symmetries under consideration~\cite{BBCDerekArx2014}.
Because of these possibilities, the concepts which were successful in demonstrating bulk-edge correspondence in static systems~\cite{BBCKaneRMP2010,BBCHatsugaiPRL1993I,BBCHatsugaiPRL1993II} may not be sufficient for doing the same in periodically driven systems.
For example, the Chern number of a Floquet band may be insufficient to predict the number of edge states. Instead, new topological invariants~\cite{BBCAsbothPRB2013,BBCRunderPRX2013} and concepts such as the choosing of special time frames~\cite{BBCAsbothPRB2013,FTCAsbothPRB2012} have to be introduced in order to capture bulk-boundary correspondence in driven systems.

In many previous studies of topological states of matter using a superlattice system,
a periodic time dependence in the model Hamiltonian is often introduced
in the form of delta-kicks (i.e., modulation of the system Hamiltonian by a delta function in time)
or a periodic sudden quench of system parameters. These studies have been fruitful. As two latest examples,
it was found by us that two periodically kicked systems may have equivalent bulk topological properties~\cite{FQHHailongPRE2013} but qualitatively different edge state behaviour~\cite{BBCDerekArx2014} due to their different chiral symmetry operators.
However, the need of an external driving field being turned on and off in an extremely narrow time window during each driving period is experimentally demanding. This fact
may then pose an obstacle to experimental studies.

In this work, we investigate the aspects of Floquet bands and the associated topological phase transitions in a one-dimensional~($1$D)~tight-binding lattice, driven by an added superlattice potential that changes continuously in time.  Such a setup with continuous time dependence
will be more realizable in experiments.  This model itself is not very new \cite{DHMMonteiroPRE2008,ExpSetKolovsky2012}. As a matter of fact,
its simple version (without a phase shift in the Hamiltonian) was previously proposed in the context of quantum chaos to study quantum state transfer through a chaotic sea~\cite{DHMMonteiroPRE2008}.  Note that
in the static case, the two-dimensional~($2$D)~parent model of this system is the Harper-Aubry-Andr\'e model~\cite{SHMHarperPPSLSA1955,SHMAA1980}.  Because the Harper-Aubry-Andr\'e model
 was extensively studied in the context of quantum Hall effects~\cite{SHMKohmotoJPSJ1992}, we naturally expect its continuously driven version to be useful for the following question: how a continuous driving field may create new topological states of matter absent in the static version?  In the following we shall simply refer to the continuously driven Harper-Aubry-Andr\'e model as the continuously driven Harper model~(CDHM).

The outline of this paper is as follows. In Sec.~II we introduce some details of CDHM.  To emphasize that many dynamical systems previously studied in the context of quantum chaos can be useful for studies of topological phase transitions, we also present the dynamics of its classical limit.
Section III is the main part of this paper and it is divided into a few subsections. In Sec.~III~A we show that the Floquet spectrum of CDHM with an odd number of bands is grouped into well-gapped bands with nonzero Chern numbers defined with respect to a Bloch phase~[under periodic boundary conditions (PBC)]~and a periodic phase-shift parameter $\beta$. The number of bands may be tuned according to our choice of the system commensurability parameter $\alpha$. As system parameters are tuned, topological phase transitions occur, which are manifested by the jumps in the band Chern numbers. In Sec.~III~B, the physical meaning of the band Chern numbers is explained through an adiabatic transport protocol. In Sec.~III~C
we show that under certain choices of the system parameters, very flat Floquet bands may emerge. The flatness of such bands is also investigated by wavepacket dynamics simulations.
In the same subsection, we show that if the number of Floquet bands is changed to be even, it is possible for the Floquet spectrum to have Dirac cones intersecting at quasienergy zero. A wavepacket prepared on such a Dirac cone is shown to move in the lattice non-dispersively.  Finally, we consider open boundary conditions~(OBC) in Sec.~III~D. There, for cases with an odd number of bands, the Floquet bulk band topology is examined in connection with the chiral edge modes traversing the bulk gap, with anomalous edge modes with the same quasienergy but opposite chiralities found on the same edge of the system. For cases with an even number of bands, interesting zero quasienergy modes are observed within the quasienergy gaps.
In Sec.~IV we discuss possible experimental realizations of the CDHM. Section V summarizes our main results and proposes future directions.

\section{A Continuously Driven Harper Model}

The classical Hamiltonian of the CDHM reads
\begin{equation}
H = J \cos(k) + V \cos(2 \pi \alpha x - \beta) \cos(\Omega t), 
\end{equation}
where $x$ and $k$ are continuous coordinate and quasi-momentum variables respectively, $\beta \in [0,2\pi)$ is a phase shift parameter \cite{motivator} and $\Omega$ is the driving frequency. The dynamical behaviour of the system is governed by its Hamilton's equations of motion
\begin{equation}
\begin{split}
\dot{x} &= -J \text{sin}(k) \qquad\qquad\qquad x \in [0,q),\\
\dot{k} &= 2 \pi \alpha V \text{sin}(2 \pi \alpha x - \beta)\cos(\Omega t) \quad k \in [0,2\pi).
\end{split}
\end{equation}
We choose $\alpha = p/q$ with $p$, $q$ being co-prime integers. Classical phase space plots are obtained by recording the values of $(x,k)$ at integer multiples of the driving period $t = n T$, with $n$ being positive integers and $T = 2\pi/\Omega$ being the driving period. Four examples of the phase space diagrams for a fixed value of $\beta$ are shown in Fig.~$1$. There we set $J = -1, V = 2, \alpha = 1/100, \beta = 0, T = 2\pi/0.7, 2\pi/0.2, 2\pi/0.16$ and $2\pi/0.12$. In the previous work ~\cite{DHMMonteiroPRE2008} without the $\beta$ parameter, it was pointed out that with the increasing of driving period $T$, the phase space plots become increasingly chaotic. This is also seen here with an arbitrary fixed value of $\beta$. From Fig.~$1$(c) and~(d) we see that all principle tori have disappeared at $T = 2\pi/0.16$. But at $T = 2\pi/0.12$, shearless tori emerge in the chaotic sea, separating it into different regions. These shearless tori have been shown to assist non-dispersive transmission of localized wavepackets in the quantized system. Here we shall focus on the topological properties of this model in the quantum case. As seen below, non-dispersive wave packet transmission can be realized due to the presence of Dirac cones in the system. This constitutes another approach for realizing localized wavepacket transport.

\begin{figure}
\begin{center}$
\begin{array}{cc}
\includegraphics[trim=0.5cm 0.9cm 0.5cm 8cm, clip=true, height=!,width=15cm] {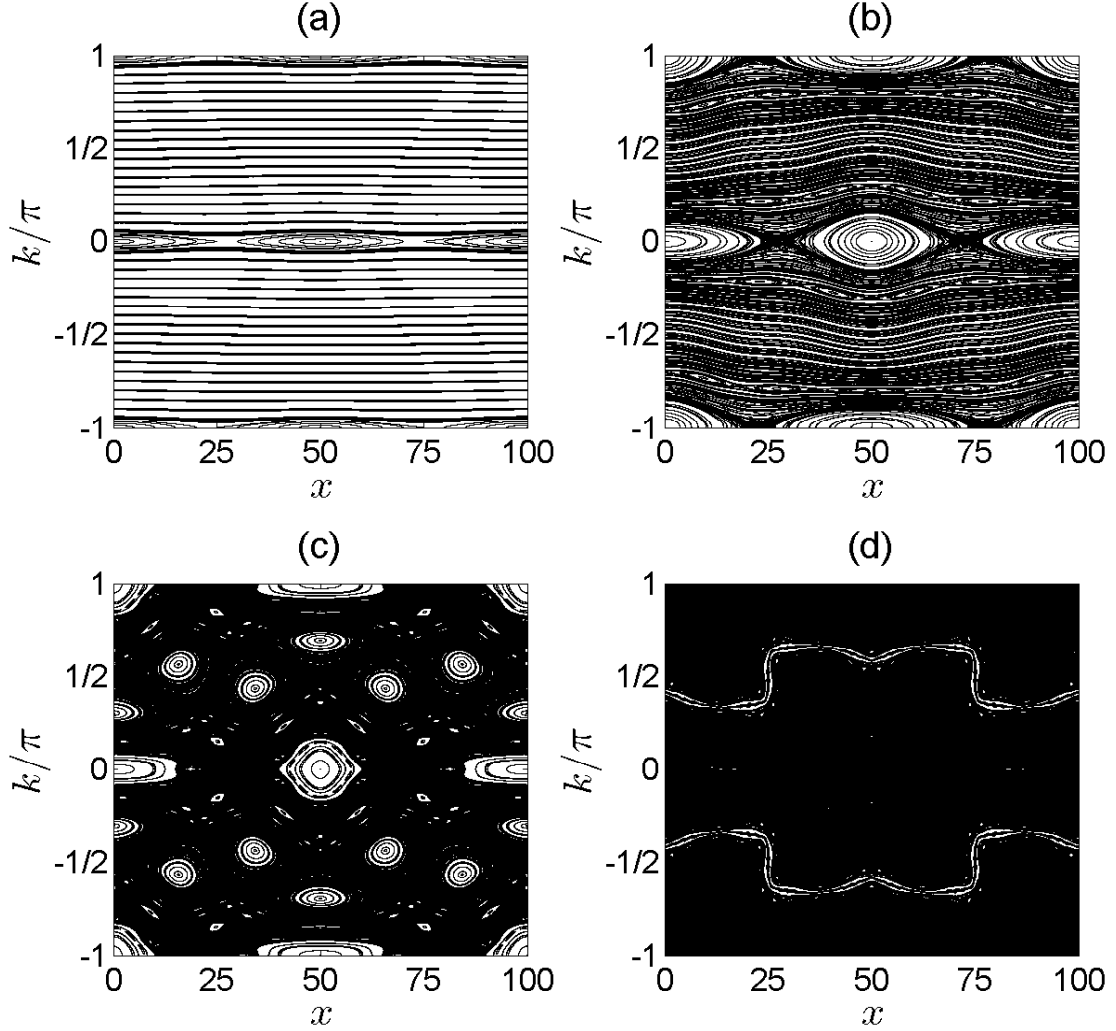}
\end{array}$
\end{center}
\caption{(color online). Classical phase space structure of the continuously driven Harper model with a fixed value of $\beta$, for (a) $T = 2\pi/0.7$, (b) $T = 2\pi/0.2$, (c) $T = 2\pi/0.16$, (d) $T = 2\pi/0.12$. The phase space becomes increasingly chaotic with the increase of the driving period. At $T = 2\pi/0.12$ the shearless tori emerge and separate the chaotic sea into different regions.} 
\end{figure}

Before discussing the quantized version of the above model, we first discuss a static version of the above model, by considering a particle hopping on a static tight-binding lattice:
\begin{equation}
\hat{H} = \sum_{m} \left[ \frac{J}{2} (a_{m}^{\dag} a_{m+1} + h.c.) + V \text{cos}(2 \pi \alpha m - \beta) a_{m}^{\dag} a_{m} \right] \hspace{0.5cm} m \in Z, \beta \in [0,2\pi), 
\end{equation}
where $J$ is the nearest-neighbor hopping amplitude and $V$ controls the strength of the superlattice potential. In this case, the lattice coordinate can only be integers. Thus the phase shift $\beta$ cannot be removed by any gauge transformation and it may be regarded as the quasi-momentum along a second dimension~\cite{FQHDerekPRL2012}. For each fixed $\beta$, the model describes non-interacting particles hopping on a $1$D lattice with a superlattice potential. In experiments, this model has been simulated using a $1$D array of evanescently coupled waveguides~\cite{SHMKraus2012} and a non-interacting BEC in a $1$D quasi-periodic optical lattice~\cite{ExpSetRoati2008}. Under the following parameter choices~\cite{SHMChenSPRL2012}:
\begin{equation}
\frac{J}{2} \Leftrightarrow J_{x} \hspace{1cm} V \Leftrightarrow 2J_{y} \hspace{1cm} \beta \Leftrightarrow k_{y} \hspace{1cm} a_{m}^{\dag} \Leftrightarrow a_{m,k_{y}}^{\dag}, 
\end{equation}
this model could be mapped onto a $2$D ``parent" model describing non-interacting electrons hopping in a square lattice subjected to a magnetic field perpendicular to the plane:
\begin{equation}
\hat{H} = \sum_{m,k_{y}} \left[ J_{x} (a_{m,k_{y}}^{\dag} a_{m+1,k_{y}} + h.c.) + 2 J_{y} \text{cos}(2 \pi \alpha m - k_{y}) a_{m,k_{y}}^{\dag} a_{m,k_{y}} \right] \hspace{0.5cm} m \in Z, k_{y} \in [0,2\pi), 
\end{equation}
where we have chosen the Landau gauge so that the electromagnetic vector potential ${\bf A} = (0,Bm,0)$. $J_{x}$ and $J_{y}$ refer to the hopping amplitudes along $x$ and $y$ directions , respectively; $m$ is the lattice index along $x$-direction and $k_{y}$ is the quasi-momentum along $y$ direction. $a_{m,k_{y}}^{\dag}$ is the creation operator for an electron in the lattice site $m$ along $x$ with quasi-momentum $k_{y}$ along $y$, and $\alpha$ represents the number of magnetic flux quanta per unit cell. This is nothing but the famous Harper-Aubry-Andr\'e model~\cite{SHMAA1980,SHMHarperPPSLSA1955}. When $\alpha$ is a rational fraction $p/q$ with $p, q$ being co-prime integers, the energy spectrum of the system takes the form of $q$ bands under PBC where each band has a nonzero Chern number~\cite{SHMKohmotoJPSJ1992}. When $\alpha$ is an irrational number, the spectrum of the system shows self-similar structures. Scanning the spectrum with respect to $\alpha$ then leads to a fractal-like spectrum called the Hofstadter butterfly~\cite{SHMHofstadter1976}.

Consider now our CDHM as a modification or a continuously driven version of the Harper-Aubry-Andr\'e model, where the superlattice potential is smoothly modulated in time by an external field. The Hamiltonian of CDHM has the following form:
\begin{equation}
\hat{H} = \sum_{m} \left[ \frac{J}{2} (a_{m}^{\dag} a_{m+1} + h.c.) + V \text{cos}(2 \pi \alpha m - \beta) \text{cos}(\Omega t) a_{m}^{\dag} a_{m} \right] \hspace{0.5cm} m \in Z, \beta \in [0,2\pi), 
\end{equation}
where $\Omega = 2\pi/T$ is the frequency of the driving field, $T$ is the driving period. Just as in the static case, the phase shift parameter $\beta$, which can be controlled externally, can be directly understood as the role of the quasi-momentum $k_y$ in a $2$D ``parent" model.   Because the Hamiltonian is time-periodic, this model can be treated within the framework of Floquet theory. In the next section it will be shown that under PBC, the Floquet spectrum is grouped into bands with nontrivial topology, which is however different from that found in the
static Harper-Aubry-Andr\'e model and in the kicked Harper model~\cite{FQHDerekPRL2012}.

\section{Floquet Band Topology}

\subsection{Chern Number Zoo}

\begin{figure}
\begin{center}$
\begin{array}{cc}
\includegraphics[trim=0.5cm 0.5cm 0.3cm 8.2cm, clip=true, height=!,width=11cm] {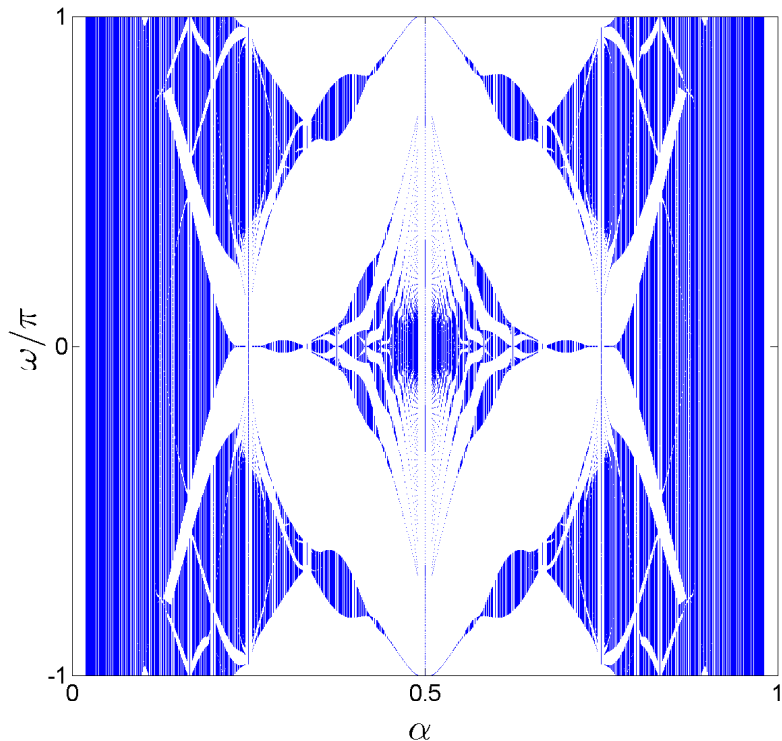}
\end{array}$
\end{center}
\caption{(color online). Butterfly Spectrum of CDHM with respect to $\alpha = p/q$, $T = 2$, $J = 2\pi/3$, $V = 2\pi$, $q$ goes from $2$ to $50$. For a better visualization of the butterfly spectrum,  the spectra for phase shift $\beta = \pi/2$ and $3\pi/4$ are superposed together.}
\end{figure}

To study topological properties of CDHM, we start with its Floquet operator~(i.e., unitary time propagator over one driving period)
\begin{equation}
\hat{U}_{\text{CDHM}}(T,0) = \hat{{\bf T}}e^{-i\int^{T}_{0}dt [J \text{cos}({\hat k}) + V \text{cos}(2 \pi \alpha {\hat m} - \beta) \text{cos}(\Omega t)]},
\end{equation}
where $\hat{{\bf T}}$ means time ordering. We have defined our system on a discrete $1$D lattice so the eigenvalues of the quasi-momentum operator $\hat{k}$ belong to $[0,2\pi)$ and the eigenvalues of the position operator $\hat{m}$ belong to integers from $-\infty$ to $\infty$. The Planck constant is set to $1$ in this work. In our study, only cases with a rational $\alpha = p/q$ are considered, where $p$ and $q$ are co-prime integers. Under the PBC, the Floquet operator commutes with the translational operator $\hat{T}_k = e^{-iq\hat{k}}$.
These two operators comprise a complete set of commuting operators whose common eigenvectors span the Hilbert space of interest for each fixed $\beta$. The spectrum of $\hat{U}_{\text{CDHM}}(T,0)$ is obtained by solving the eigenvalue problem $\hat{U}_{\text{CDHM}}(T,0) | \psi \rangle = e^{-i \omega} | \psi \rangle $,
where $\omega$ is referred to as the eigenphase. The quasienergy is the eigenphase divides by the driving period. Using the well-known split operator method, we numerically solve for $\omega$  with fixed $T$, $J$ and $V$ values at all points in the $2$D  Brillouin zone (BZ) defined by $\beta$ and a Bloch phase $\phi$~[i.e., the eigenphase of the translational operator $\hat{T}_k$ which belongs to $[0,2\pi)$]. Due to the translational symmetry of $\hat{U}_{\text{CDHM}}(T,0)$, the eigenphases come in $q$ bands which we refer to collectively as the Floquet spectrum. In Fig.~$2$, we show the eigenphases of the CDHM as a function of $\alpha = p/q$ where $q$ goes from $2$ to $50$, $p$ takes integer value from $1$ to $q-1$ for each $q$ and $\phi$ is scanned from $0$ to $2\pi$ for each $\alpha$. The spectrum has a fractal-like structure similar to the ones previously observed in two kicked models with nontrivial band topology~\cite{FQHGongPRA2008}. In Fig.~$3$, we show several examples of the Floquet spectrum where we choose $T = 2$ and $\alpha = 1/3$ or $1/5$. With changes in $J, V$ and $T$, the bands may deform, touch and re-separate, accompanied with possible topological phase transitions. The topological properties of a Floquet band well separated from the others can be characterized by its Chern number~\cite{FQHDerekPRL2012}:
\begin{equation}
C_{n} = \frac{i}{2\pi} \oint d{\bf k} \cdot \bra{\psi_{n}(\phi,\beta)} \nabla_{{\bf k}} \ket{\psi_{n}(\phi,\beta)} \hspace{1cm} {\bf k} = (\phi,\beta),
\end{equation}
where $n$ is the band index and $\ket{\psi_{n}(\phi,\beta)}$ is an eigenstate on band $n$ for a given Bloch phase and phase shift. These values may change~(i.e., a topological phase transition occurs)~only when the Floquet bands meet at some points in the Brillouin zone. Numerically, this quantity can be calculated using the standard method in Ref.~\cite{RestaNotes2013}.

\begin{figure}
\begin{center}$
\begin{array}{cc}
\includegraphics[trim=0.5cm 0.9cm 0.5cm 8cm, clip=true, height=!,width=15cm] {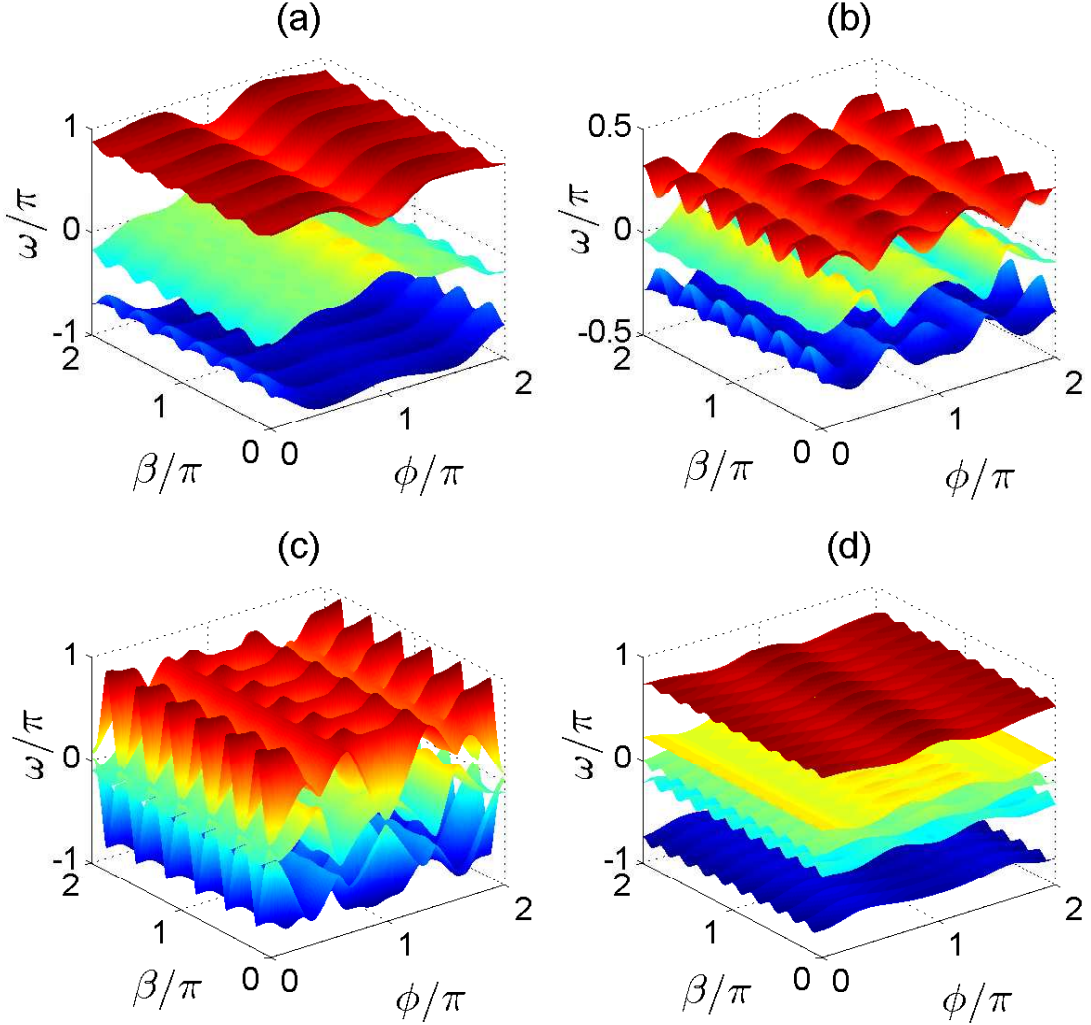}
\end{array}$
\end{center}
\caption{(color online). The eigenphase spectrum of CDHM for (a) $T = 2, \alpha = 1/3, J = V = 4.1$, the three bands have Chern numbers $4, -8, 4$~(from bottom to top), (b) $T = 2, \alpha = 1/3, J = V = 5.4$, the three bands have Chern numbers $-8, 16, -8$~(from bottom to top), (c) $T = 8.4, \alpha = 1/3, J = 3.0, V = 4.0$, the three bands have Chern numbers $16, -32, 16$~(from bottom to top), (d) $T = 2, \alpha = 1/5, J = V = 6.3$, the five bands have Chern numbers $-4, 6, -4, 6, -4$~(from bottom to top).} 
\end{figure}

In Table~$1$ and $2$ we show the band Chern numbers of the above introduced CDHM with $T = 2$ and $\alpha = 1/3, 1/5$. For simplicity, we choose $J = V$ and scan $J$ from $0.1$ to $10$ for the $3$-band case and from $2.0$ to $6.9$ for the $5$-band case, in steps of $0.1$. $C_1$, $C_2$ and $C_3$ refer to the Chern numbers of the three bands~(from bottom to top)~respectively in the $3$-band case. In the five band case, they refer to the first three bands~(from bottom to top). We omit the Chern numbers of bands $4$ and band $5$ because they are the same as those of bands $2$ and $1$ for all the parameters shown here.

\begin{table}[htbp]
\caption{Chern numbers of CDHM with respect to $J$ and $V$ for $\alpha = 1/3, T = 2$}
\vskip 1em
\centering
\begin{tabular}{|c|c|c|c|c|c|c|c|}\hline
Values of $J$ and $V$ & $[0.1,3.4]$ & $[3.5,5.1]$ & $[5.2,5.4]$ & $[5.5,5.6]$ & $5.7$ & $[5.8,7.5]$ & $[7.6,10]$\\[0.5ex]
\hline
$C_1$ & $-2$ & $4$ & $-8$ & $-2$ & $-8$ & $4$ & $-2$ \\
\hline
$C_2$ & $4$ & $-8$ & $16$ & $4$ & $16$ & $-8$ & $4$ \\
\hline
$C_3$ & $-2$ & $4$ & $-8$ & $-2$ & $-8$ & $4$ & $-2$ \\
\hline
\end{tabular}
\end{table}

\begin{table}[htbp]
\caption{Chern numbers of CDHM with respect to $J$ and $V$ for $\alpha = 1/5, T = 2$}
\vskip 1em
\centering
\begin{tabular}{|c|c|c|c|c|c|c|c|c|c|}\hline
Values of $J$ and $V$ & $[2.0,3.7]$ & $[3.8,3.9]$ & $[4.0,4.2]$ & $[4.3,4.9]$ & $[5.0,5.5]$ & $[5.6,6.6]$ & $6.7$ & $6.8$ & $6.9$ \\[0.5ex]
\hline
$C_1$ & $6$ & $-4$ & $-4$ & $6$ & $6$ & $-4$ & $-4$ & $-4$ & $-4$ \\
\hline
$C_2$ & $-4$ & $6$ & $-4$ & $-14$ & $-4$ & $6$ & $-4$ & $26$ & $-4$ \\
\hline
$C_3$ & $-4$ & $-4$ & $16$ & $16$ & $-4$ & $-4$ & $16$ & $-44$ & $16$ \\
\hline
\end{tabular}
\end{table}

\begin{table}[htbp]
\caption{Chern numbers of CDHM with respect to $T$ for $\alpha = 1/3, J = 3.0, V = 4.0$}
\vskip 1em
\centering
\begin{tabular}{|c|c|c|c|c|c|c|c|c|c|}\hline
Values of $T$ & $[4.0,4.7]$ & $[4.8,5.8]$ & $[5.9,6.5]$ & 6.6 & $[6.7,7.1]$ & $[7.2,8.0]$ & $[8.1,8.3]$ & $[8.4,8.9]$ & $9.0$ \\[0.5ex]
\hline
$C_1$ & $4$ & $-2$ & $4$ & $-8$ & $-2$ & $4$ & $-2$ & $16$ & $4$ \\
\hline
$C_2$ & $-8$ & $4$ & $-8$ & $16$ & $4$ & $-8$ & $4$ & $-32$ & $-8$ \\
\hline
$C_3$ & $4$ & $-2$ & $4$ & $-8$ & $-2$ & $4$ & $-2$ & $16$ & $4$ \\
\hline
\end{tabular}
\end{table}

For a fixed pair of $J$ and $V$, topological phase transitions can also be induced via changing the driving period $T$. The Chern numbers in a $3$-band case with $J = 3.0, V = 4.0$ and $T$ going from $4.0$ to $9.0$ (with a step size of $0.1$) are given in Table~3. Several interesting observations can be made from the Chern number zoo. Firstly, with the change of $J$, $V$ and $T$, many topological phase transitions occur. Secondly, Floquet bands with very large Chern numbers emerge at certain parameter values. Finally, in all the cases considered here, the Chern number of every Floquet band is an even integer (this feature will be explained in Appendix B). Compared with the static case in which all bands except the middle one have a Chern number $+1$ and no topological phase transitions will be induced via varying $J$ and $V$ for a fixed $\alpha$, it is clear that the periodic driving has introduced some novel features absent in the Harper-Aubry-Andr\'e model.
The Chern number zoo shown here is also much different from what was found in the kicked Harper model~\cite{FQHHailongPRE2013}. These observations confirm that, in addition to the well-studied kicked-Harper model (quantized on a torus or on a cylinder)~\cite{FQHHailongPRE2013,leubolf,dana}, our CDHM is indeed rich enough for the study of topological states in driven systems.

\subsection{Quantized Adiabatic Pumping}

The nontrivial Floquet band topology is evidenced physically through quantized adiabatic pumping. Based on previous work~\cite{TPumpThoulessPRB1983,FQHDerekPRL2012}, we expect and indeed verify that a band Wannier state, formed by uniformly superposing all eigenstates of a Floquet band with a nonzero Chern number, will move over an integer number of lattice sites under an adiabatic change of $\beta$ from $0$ to $2\pi$. The change of the wavepacket center during this process is given by the Chern number of the band associated with the Wannier state:
\begin{equation}
\langle\hat{m}\rangle = \bra{W_{n}(NT)}\hat{m}{\ket{W_{n}(NT)}} - \bra{W_{n}(0)}\hat{m}\ket{W_{n}(0)} = q C_{n}
\end{equation}
where $\ket{W_{n}(0)}$ and $\ket{W_{n}(NT)}$ refer to the Wannier states constructed from band $n$ at the initial and final times of the adiabatic cycle, $T$ is the driving period, $N$ is the total number of driving periods used to complete the cycle and $\hat{m}$ is the discretized position operator. In our system, a Wannier state on a Floquet band can be prepared following the methods in~\cite{FQHDerekPRL2012} and the periodic parameter for the adiabatic cycle can be chosen as the phase shift $\beta$, which takes the same value within each driving period and changes slowly after each $T$. As an example, in Fig.~$4$(b) and (d) we show the Wannier state evolution during an adiabatic cycle for the $3$-band and $5$-band cases. The results show that the change of the wavepacket center in this adiabatic cycle is indeed equal to the Chern number of the related band. The accuracy of the quantization improves with the increase of the total number of driving periods within each adiabatic cycle.

\begin{figure}
\begin{center}$
\begin{array}{cc}
\includegraphics[trim=0.5cm 0.7cm 0.3cm 8cm, clip=true, height=!,width=15cm] {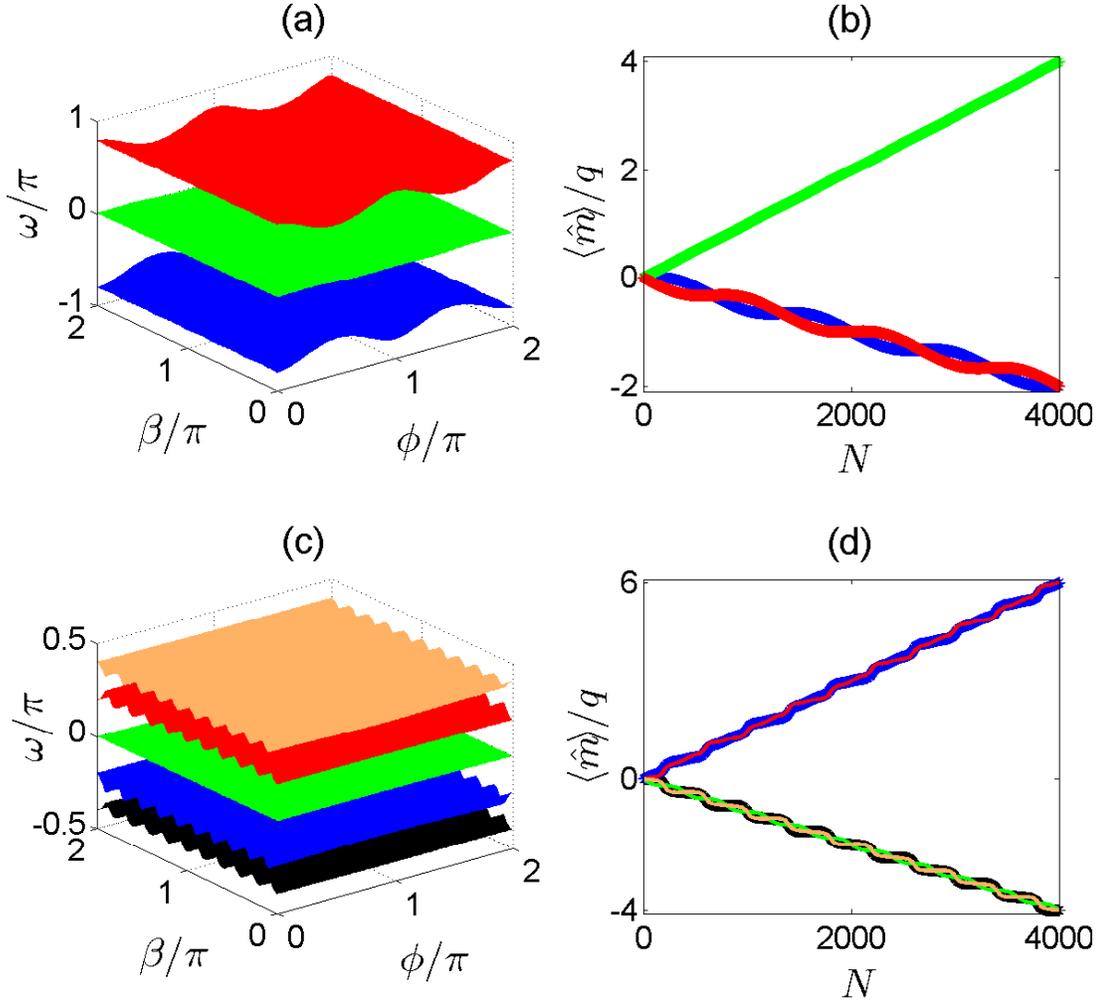}
\end{array}$
\end{center}
\caption{(color online). Quantized pumping in the position space for $T = 2$. (a) Eigenphase spectrum for $\alpha = 1/3, J = V = 2.5$, the three bands have Chern numbers $-2, 4, -2$~(from bottom to top). (b) Adiabatic pumping results for Wannier states prepared from the $3$ bands in (a). The adiabatic cycle is completed in $4000$ driving periods and the difference $\langle\hat{m}\rangle$ of the position expectation value is taken at $t = NT$ for $N = 1, 2, ..., 4000$. The color of the bands and the pumping curves are matched with each other. For all the three cases, the difference between the wave packet center at the end and the beginning of the adiabatic cycle equals the band Chern numbers. (c) Eigenphase spectrum for $\alpha = 1/5, J = 2\pi/5, V = 2.4\pi$. The five bands have Chern numbers $-4, 6, -4, 6, -4$ (from bottom to top). (d) Adiabatic pumping results for Wannier states prepared from the $5$ bands in (c). For all the five cases, the difference between the wave packet center at the ending and the starting point of the adiabatic cycle equals the band Chern numbers.} 
\end{figure}

At topological phase transition points, the band gap  between two bands vanishes, leading to the breakdown of the adiabatic condition for the evolution of Wannier states on the bands. Thus we should expect a breakdown of the quantized adiabatic pumping around the phase transition point. This is shown in Fig.~$5$ where we can see that~(i)~away from the phase transition point, the pumping results show good quantization behavior,~(ii)~around the phase transition point, some irregular jumps are observed in the pumping results.
If one could experimentally prepare a Wannier state on a Floquet band, the quantized adiabatic pumping might be used to detect topological phase transitions in CDHM. Admittedly, this is a challenging task as compared with the static case.

\begin{figure}
\begin{center}$
\begin{array}{cc}
\includegraphics[trim=0.5cm 0.5cm 0.3cm 8.5cm, clip=true, height=!,width=11.5cm] {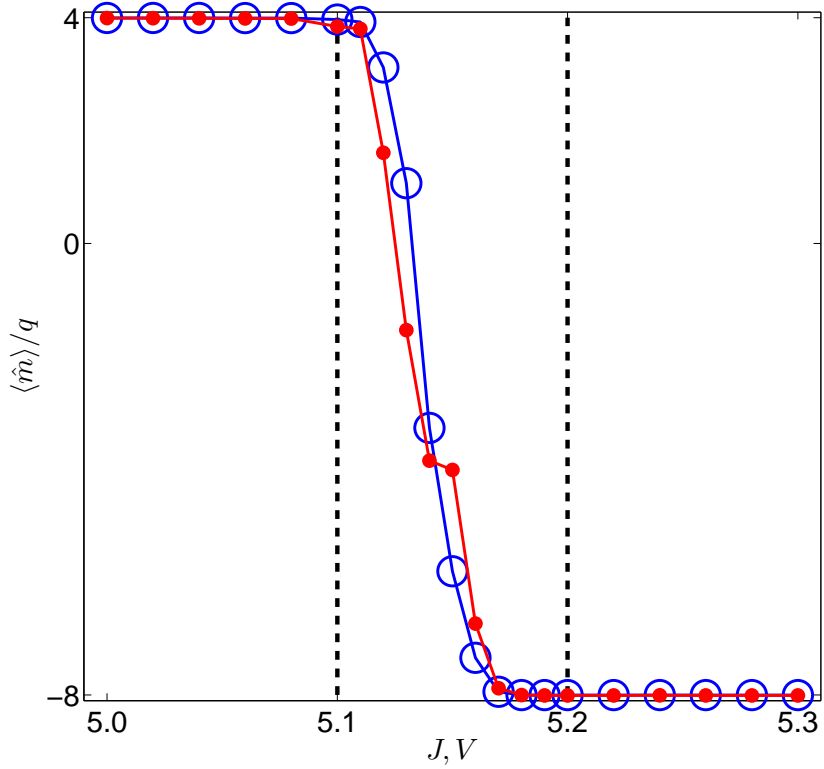}
\end{array}$
\end{center}
\caption{(color online). Quantized Pumping across a phase transition point for $\alpha = 1/3$ and $T = 2$. The difference $\langle\hat{m}\rangle$ between the final and initial position expectation values of a Wannier state initially prepared on the top band with $\beta = 0$ is shown for adiabatic cycles comprising $3000$~(red dots)~and $4000$~(blue circles)~driving periods. The phase transition happens in the domain between the black dashed lines~(from $J = V = 5.1$ to $J = V = 5.2$). At $J = V = 5.1$, the top band has a Chern number $+4$. At $J = V = 5.2$, the top band has a Chern number $-8$.}
\end{figure}

\subsection{Flat Bands and Dirac Cones}

Flat bands~\cite{FBLiuIJOMPB2013,FBSondhiCRP2013,FBWuArx2014} are bands in which the energy spectrum is dispersionless with respect to the crystal momentum. On such bands, particles will have arbitrarily large effective mass and their kinetic energies are quenched. The physical properties of the system will be pre-dominantly determined by particle-particle interactions. If the flat band has a nonzero Chern number, many-body interactions and band topology may interplay and lead to the emergence of exotic  states of matter~(e.g., the fractional Chern insulators--fractional quantum Hall effects without Landau levels). Another interesting feature associated with band spectra is emergence of Dirac cones. These are regions in the band spectrum where two bulk bands meet in a conical intersection. In this case, states on the linear-dispersion part of the band will have zero effective mass and thus transport non-dispersively in the system.
Dirac-cone physics has been extensively studied in the context of condensed matter physics and one of its famous realizations is that of graphene~\cite{WolfBook2014}. In this subsection we will see that the manifestation of both flat bands and Dirac cones in Floquet systems may be studied in CDHM, a fact that may motivate the simulation of Dirac-cone physics and flat-band physics with continuously driven 1D systems.

\begin{figure}
\begin{center}$
\begin{array}{cc}
\includegraphics[trim=0.5cm 0.5cm 0.3cm 10cm, clip=true, height=!,width=15cm] {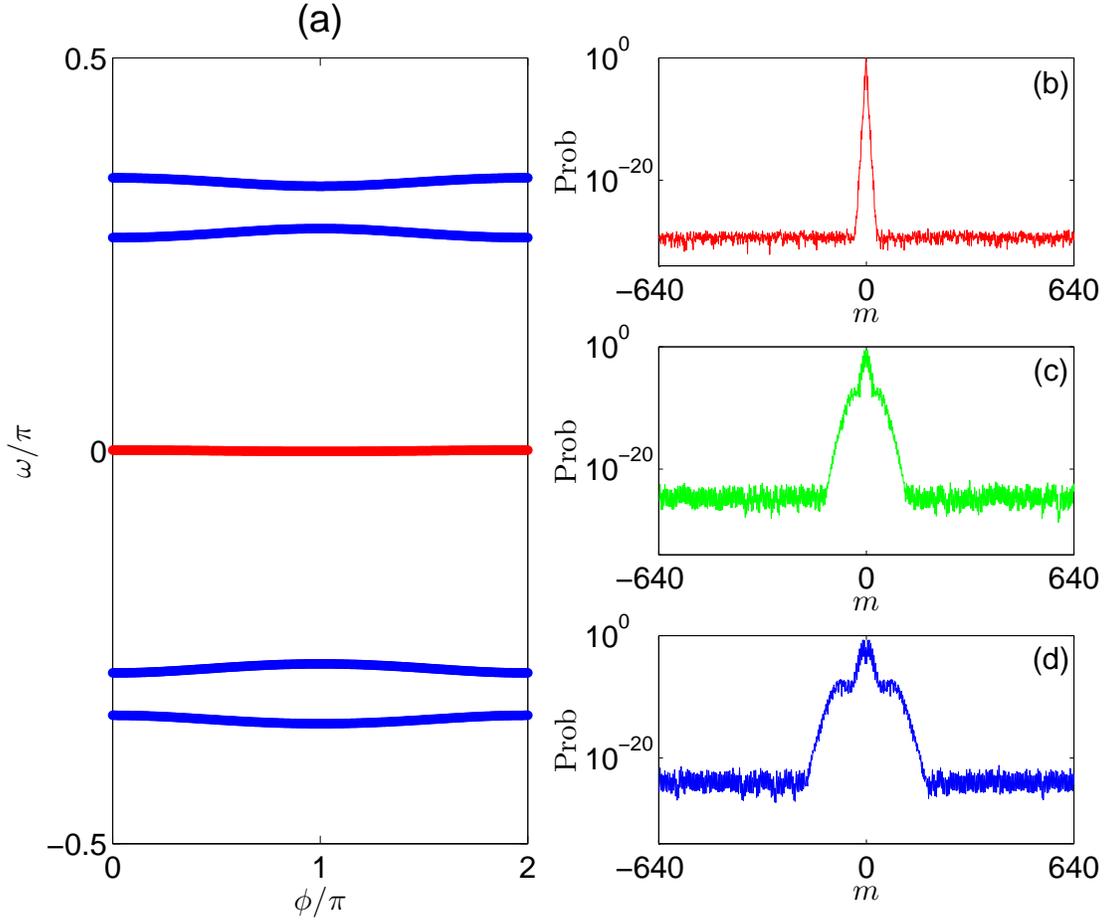}
\end{array}$
\end{center}
\caption{(color online). Evolution of a Wannier state for $T = 2, \alpha = 1/5, J = 2\pi/5, V = 2.4\pi, \beta = 0.1\pi$ on a flat band. (a) Eigenphase with respect to the Bloch phase $\phi$. Red line: the middle nearly-flat band, blue line: other non-flat bands. (b)-(d) evolution of a Wannier state prepared on the middle band at $t = 0$. $x$-axis: lattice index $m$ which goes from $-640$ to $640$. $y$-axis: the probability distribution in the lattice of (b) the initial state, (c) the state at $t = 500T$ and (d) the state at $t = 1000T$ in a base $10$ logarithmic scale. It is clear that the Wannier state initially prepared on the flat band spreads very slowly in the $1$D lattice.} 
\end{figure}

First, in both $3$ and $5$-band cases, the middle band of CDHM changes its bending direction~(e.g., from upward to downward)~as $J$ and $V$ are varied. At certain parameter values during this process, the band may become very flat. In Fig.~$4$ two examples of flat bands for $\alpha = 1/3$ and $\alpha = 1/5$ are given. In Fig.~$4$(a) the flatness ratio, defined as the minimum direct band gap divided by band width, of the middle band is approximately $23$ while that in Fig.~$4$(b) is approximately $86$.

On a flat band, a Wannier state will spread very slowly due to the almost zero band curvature. In Fig.~$6$, we prepare a Wannier state on the middle band of Fig.~$4$(b) and evolve it in our $1$D superlattice. The results show that even after $1000$ driving periods, the state is still well localized around its original center. So by filling several particles on this flat band, their interactions might dominate the physics of the system and the interplay between Floquet band topology and interactions may be studied~\cite{FloqIntOkaArx2013,FloqIntSantoroPRL2012,FFCINeupertPRL2014}.

 One more interesting feature of this model arises when the total number of bands is even~(i.e., $q$ equals an even integer), two Floquet bands may touch at quasienergy zero. Numerically, it is found that for cases with 4 bands (but not for cases with 2 bands), the dispersion relation in the neighborhood of the touching bands becomes linear in terms of both $\beta$ and the Bloch phase $\phi$, thus forming Dirac cones.  Remarkably, it is also found that each of these Dirac cones has a $\pi$ Berry phase. This is intriguing because in the context of graphene physics, such kind of $\pi$ Berry phase is the underlying physics responsible for the half-quantized anomalous Hall conductance.
In Fig.~$7$(a) an example of the Floquet Dirac cone is shown in a $4$-band case. Wavepackets prepared along the same side of the Dirac cone can transport non-dispersively. In Fig.~$7$(c) we show one such example of wavepacket evolution in our system. The initial state is prepared as a superposition of $20$ band eigenstates along the same side of the Dirac cone~[residing along the red line in Fig.~$7$(b)]. The simulation result clearly shows that the wavepacket evolves in time without changing its shape.

\begin{figure}
\begin{center}$
\begin{array}{cc}
\includegraphics[trim=0.5cm 0.1cm 0.3cm 9.9cm, clip=true, height=!,width=15cm] {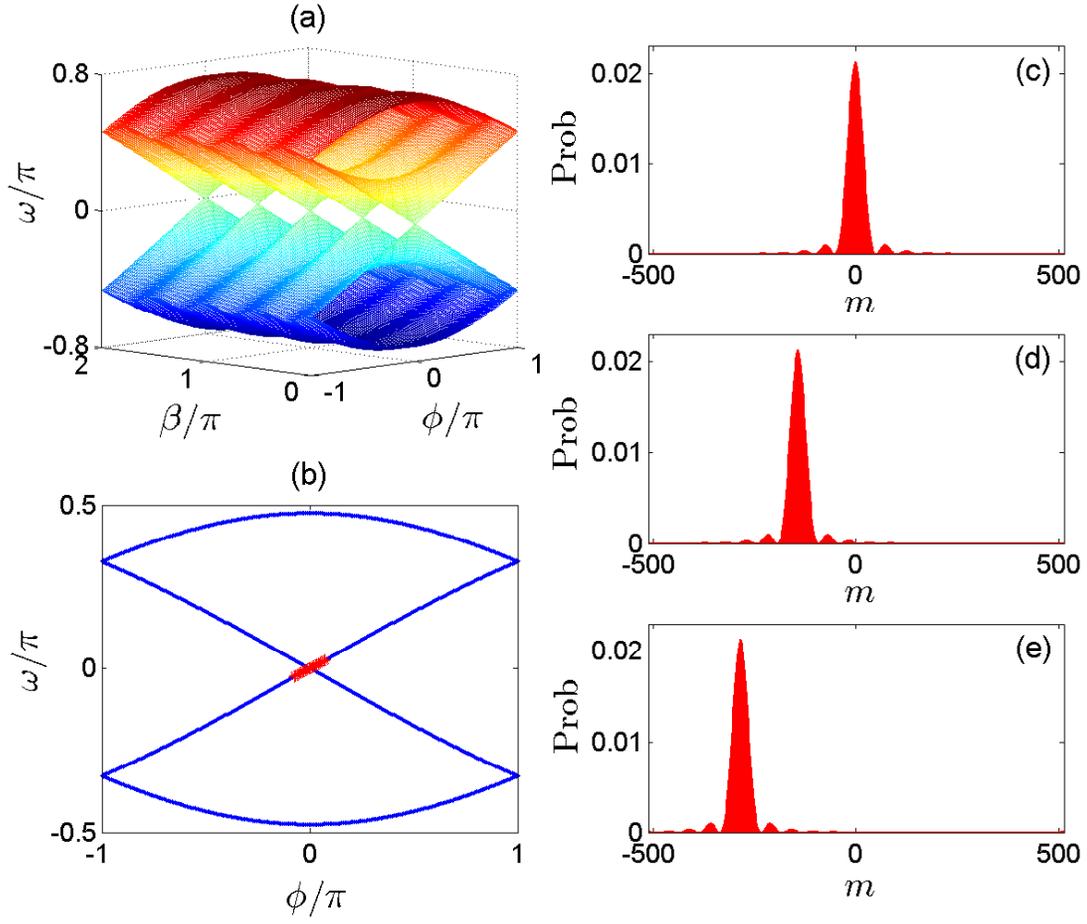}
\end{array}$
\end{center}
\caption{(color online). Dirac cones and non-dispersive transport in a 4-band case.
(a) Eigenphase spectrum for $\alpha = 1/4, J = \pi/2, V = \pi$. The Bloch phase is scanned from $-\pi$ to $\pi$ to show the Dirac cones clearly.
(b) Eigenphase spectrum on the plane $\beta = 0$. The part indexed with red stars show the states used in the pumping.
(c)-(e) Non-dispersive transport of states around the Dirac cones in a $1$D lattice. The length of the lattice is $L = 1024$, the lattice index $m$ goes from $-512$ to $511$. The initial state in (c) is a superposition of $20$ band eigenstates along one branch of the Dirac cone~(indexed by red stars) in (b). After $M = 100$ driving periods it evolves to the state shown in (d) and after $M = 200$ driving periods it evolves into the state in (e). It is clear that the wave packet hardly changes its shape during the evolution.} 
\end{figure}

\subsection{Chiral Symmetry and Bulk Edge Correspondence}

In the previous subsection, we studied the topological property of CDHM under PBC. The principle of bulk-edge correspondence~\cite{BBCHatsugaiPRL1993I,BBCHatsugaiPRL1993II,BBCKaneRMP2010} states that there exists a relation between bulk topology under PBC and boundary modes under OBC. The bulk-edge correspondence in driven systems is more subtle than in static systems. For example, chiral edge states may be found between two bands with zero Chern numbers if a winding edge state connects them through the borders of the quasienergy BZ~\cite{BBCRunderPRX2013}. In some cases, edge modes localized around the same edge with opposite chiralities may be observed within the same quasienergy band gap~\cite{FQHZhaoErHPRL2014}. In this section, we study the bulk-edge correspondence as well as the chiral symmetry of CDHM.

\begin{figure}
\begin{center}$
\begin{array}{cc}
\includegraphics[trim=0.5cm 0.5cm 0.3cm 8.2cm, clip=true, height=!,width=15cm] {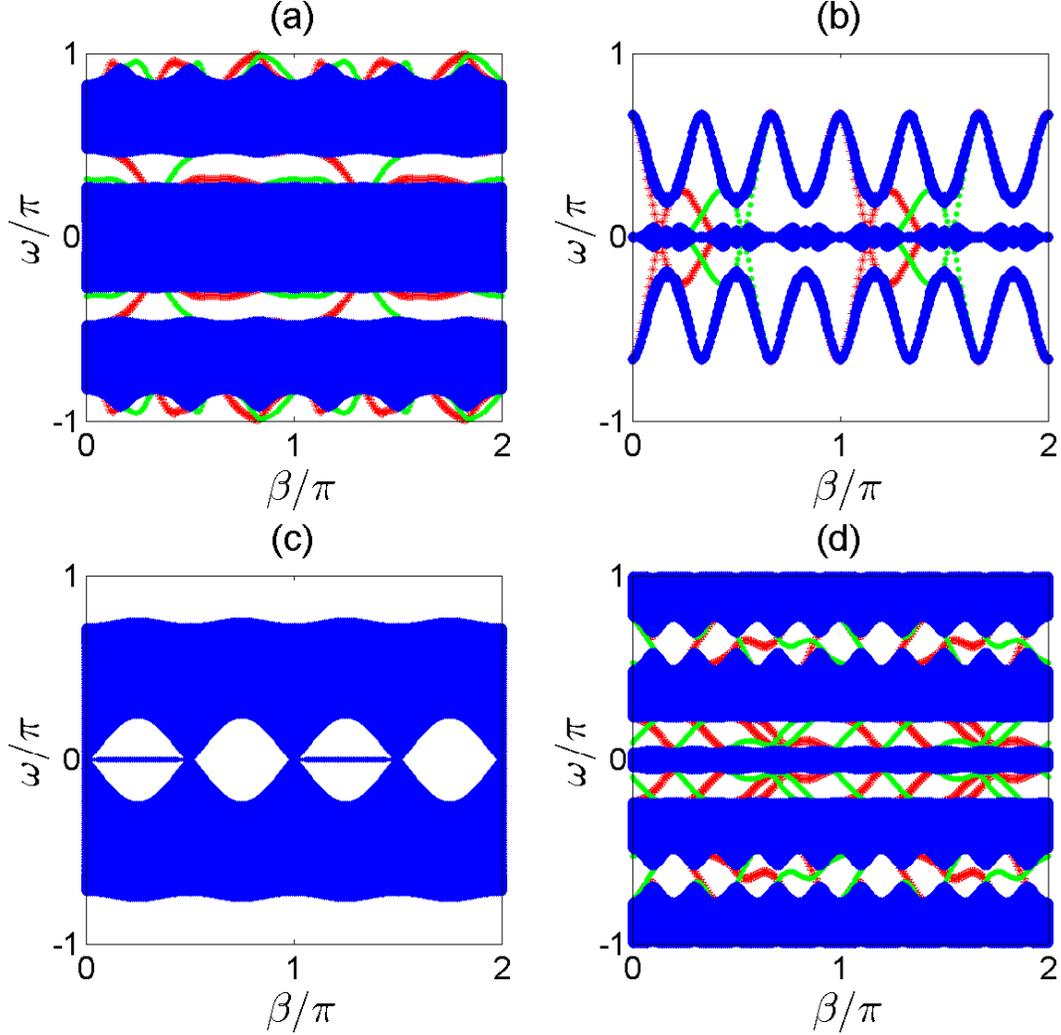}
\end{array}$
\end{center}
\caption{(color online). Eigenphase spectrum of CDHM with respect to the phase shift $\beta$ at OBC with $T = 2$ for different parameter combinations. Chiral edge modes localized on the left and right edges are indicated by red stars and green dots respectively.
When $\alpha=1/3$ with (a) $ J = V = 8.0$ and (b) $J = 2\pi/3, V = 3.2\pi$, the band Chern numbers are $-2, 4, -2$ in both cases~(from bottom to top). Anomalous counter-propagating edge modes are seen in the middle gap in panel~(b).
(c) When $\alpha = 1/4, J = \pi/2, V = \pi$, degenerate zero modes are seen in the gaps with zero eigenphase.
(d) When $\alpha = 1/5, J = V = 4.5$, the $5$ bands have Chern numbers $6, -14, 16, -14, 6$~(from bottom to top).} 
\end{figure}

A Floquet operator $\hat{U}$ with chiral symmetry means that there exists a unitary and Hermitian operator $\Gamma$ such that
\begin{equation}
\Gamma\hat{U}\Gamma^{\dagger} = \hat{U}^{-1}, 
\end{equation}
for $\hat{U}$ chosen to propagate over a special choice of one-period time interval~\cite{BBCAsbothPRB2013}. For simplicity, we consider our $1$D super-lattice with two open ends. Its Floquet operator is given by:
\begin{equation}
\hat{U}_{\text{CDHM}}(T,0)
= \hat{{\bf T}}e^{-i\int^{T}_{0}dt \sum^{L-1}_{m=1} J \left( \ket{m} \bra{m+1} + h.c. \right) + \sum^{L}_{m=1} V \text{cos}(2 \pi \alpha m - \beta) \text{cos}(\Omega t) \ket{m} \bra{m}},
\end{equation}
where $L$ is the length of the lattice. Diagonalizing $\hat{U}_{\text{CDHM}}(T,0)$ at each phase shift $\beta$ and scanning $\beta$ from $0$ to $2\pi$, we obtain the eigenphase spectrum as a function of $\beta$. We find that $\hat{U}_{\text{CDHM}}(T,0)$ obeys the chiral symmetry condition of Eq.~($10$), with the following chiral symmetry operator
\begin{equation}
\Gamma_{\text{CDHM}} = e^{i\hat{m}\pi},
\end{equation}
where $\hat{m}$ is the discrete position operator. $\Gamma_{\text{CDHM}}$ performs a local unitary transformation in the sense that it does not translate states over lattice sites. In Appendix A, we prove that due to the chiral symmetry, the spectrum of $\hat{U}_{\text{CDHM}}(T,0)$ possesses mirror symmetry with respect to zero quasienergy for each $\beta$ and is $\pi$-periodic in $\beta$.

In Fig.~$8$ we show four examples of the spectrum of $\hat{U}_{\text{CDHM}}(T,0)$ under OBC. In Fig.~$8$(a), the edge modes are correctly predicted by the Chern numbers in the same way as in the integer quantum Hall effect~\cite{BBCHatsugaiPRL1993I,BBCHatsugaiPRL1993II,BBCRunderPRX2013}. Namely, the Chern number of each band equals the difference between the number of chiral modes~(its sign in accord to their chirality)~above and below the band on a single edge of the system. For example, considering the right edge of the system considered in Fig.~~$8$(a), the middle band has two chiral modes in the gap below it with negative group velocities and two chiral modes above it with positive group velocities~[Here the group velocity is defined by interpreting $\beta$ as another quasi-momentum. It acquires its exact physical meaning through the mapping in Eq.~($4$)].
Counting each chiral mode with positive~(negative)~group velocity as $+1\,(-1)$, the bulk-edge correspondence is clear from the fact that the difference between the chiral modes above and below the band is equal to the band's Chern number~[i.e., $C_2 = +4 = +2 - (-2)$]. Similar analysis applies for the other bands.  As such, in this particular case the knowledge of all the band Chern numbers does determine the number of edge modes in the system within each gap in this case. This is no longer true for the case in Fig.~$8$(b)for a different set of system parameters. There we see that above and below the middle band there exist chiral edge modes localized on the same edge traversing the band gap with opposite chirality. Such anomalous edge modes were first reported in an earlier study of the kicked Harper model~(or a kicked quantum Hall system)~\cite{FQHZhaoErHPRL2014}. They are anomalous in the sense that one cannot correctly predict their existence just by knowing the Chern numbers of all the bands as we just did  for the case in Fig.~$8$(a). As a final remark, we note that the existence of these anomalous edge modes could not be explained through a simple chiral symmetry analysis as in Ref.~\cite{BBCDerekArx2014}.  Our results here hence call for a more general understanding (one possible route is to extend a simple model constructed in Ref.~~\cite{FQHZhaoErHPRL2014}) of anomalous edge modes in periodically driven systems.   In Fig.~8(d), the bulk band Chern numbers are chosen to be large, and as expected, many edge modes are found and their behavior is seen to be rather complicated (but consistent with our chiral symmetry analysis). In particular, one edge mode may start to emerge from a bulk band and then ends in the same bulk band. Also interesting, one edge mode across a band gap may also have opposite group velocities at different values of $\beta$ [see the edge mode in the top gap in Fig.~8(d)]. These characteristics can be understood as deformation of edge modes \cite{Enotes}, but an analysis regarding under what conditions this can happen is beyond the scope of this work.


In Fig.~$8$(c), we show the spectrum of CHDM in a $4$-band case where degenerate zero modes are observed in the spectral gap.  Because the emergence of such zero modes is also found in other kicked models, this is not a totally unexpected result. However, once again our observation here indicates that a continuously driven model can possess as many interesting aspects as in the kicked Harper model, with its own features as well. It would be interesting to consider possible detection methods of these zero modes because these degenerate zero modes have potential applications in quantum information.

\section{Possible Experimental Realizations}

To consider an experimental realization of our model, we first rewrite time-dependent Hamiltonain as the following (see also Appendix B),
\begin{eqnarray}
\hat{H}_{\text{I}} &= & \sum_{m} \left[ \frac {J}{2} \left( \ket{m}\bra{m+1} + h.c. \right)\right]
 \nonumber \\
 &+& \sum_{m}\left[ \frac {V}{2} \ket{m}\bra{m}\cos(2 \pi \alpha m + \Omega t)\right] \nonumber \\
 & +& \sum_m \left[\frac {V}{2} \ket{m}\bra{m}\cos(2 \pi \alpha m - 2\beta - \Omega t) \right].
\end{eqnarray}
Here $\ket{m}$ refers to the Wannier state localized at lattice site $m$, $\alpha = k_2/k_1$ is the ratio of two lattice wave numbers, $J/2$ is the nearest-neighbor tunneling energy and $V/2$ is the strength of the superlattice potentail.  When the driving frequency $\Omega = 0$, this can still be regarded as a static Harper model, but with two superlattice potentials of a relative phase shift $2\beta$. We note that the Harper model was experimentally realized in Ref.~\cite{ExpSetRoati2008} by loading a non-interacting BEC into a $1$D optical lattice accompanied by a secondary lattice potential.  Certainly,  it is also interesting to consider to experimentally realize our model using photonic setups~\cite{SHMKraus2012,BBCDerekArx2014}, but we focus on a cold-atom setup below.

 The time-dependent superlattice potential  $\sum_{m}\left[ \frac {V}{2} \ket{m}\bra{m}\cos(2 \pi \alpha m + \Omega t)\right]$ with $\Omega \neq 0$ may be realized by linearly chirping the frequencies of two counter-propagating waves ~\cite{ExpSetKolovsky2012}. This being the case, it seems not challenging to form a second analogous superlattice potential  $\sum_m \left[\frac {V}{2} \ket{m}\bra{m}\cos(2 \pi \alpha m - 2\beta - \Omega t) \right]$, shifted from the first one by a phase shift $2\beta$.

 \begin{figure}
\begin{center}$
\begin{array}{cc}
\includegraphics[trim=0.5cm 1cm 0.3cm 8.3cm, clip=true, height=!,width=15cm] {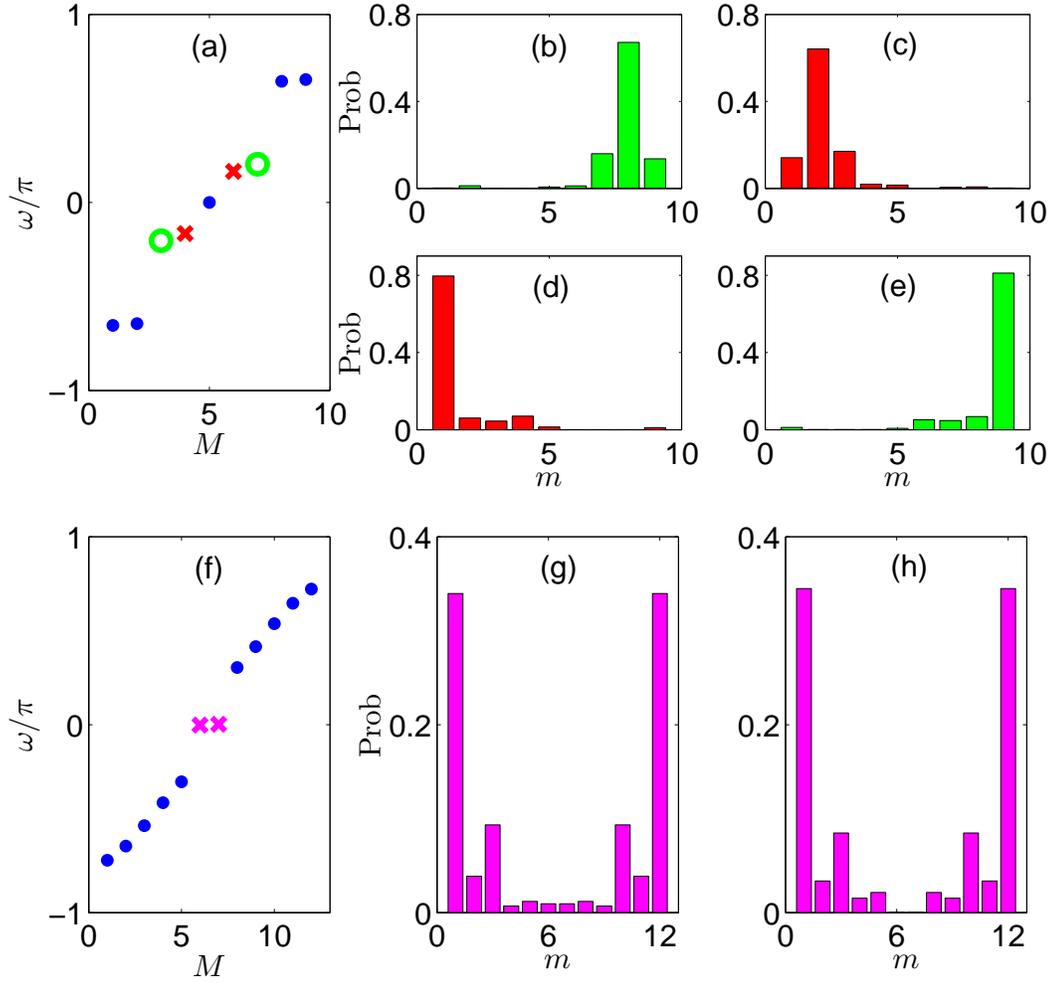}
\end{array}$
\end{center}
\caption{(color online). Edge states of CDHM in a short lattice with $T = 2$ for different parameter regimes. Bulk states are indexed with blue dots.
(a) Eigenstates of CDHM in a $3$-band case with $T = 2, J = 2\pi/3, V = \pi, \beta = 0.35\pi$. The length of the lattice is $L = 9$ while $M$ refers to the index of the eigenstate (indexed in increasing order with respect to the eigenphase). Edge states localized on the right edge ($M = 3, 7$) are indicated by green circles while those on the left edge ($M = 4, 6$) are indicated by red crosses.
(b)-(e) The probability distributions of states with index $M = 3, 4, 5, 7$ in Fig.~(a). The total occupation probabilities in the first or the last three lattice sites are approximately $0.97, 0.95, 0.90 ,0.93$ in these four cases respectively.
(f) Eigenstates of CDHM in a $4$-band case with $T = 2, J = \pi/2, V = \pi, \beta = \pi/4$. The length of the lattice is $L = 12$. Edge states ($M = 6, 7$) localized around both edges are indicated by magenta crosses.
(g)-(h) The probability distribution of states with index $M = 6, 7$ in Fig.~(f). The total occupation probability in the first and last four lattice sites are both approximately $0.96$ in the two cases.
} 
\end{figure}

In such a cold-atom setup, the wavepacket dynamics could be tracked by taking a picture of the atomic cloud after each driving period~\cite{ExpSetRoati2008}. Somewhat stimulating, we find that a $1$D lattice with very few sites (around 10 sites only) is already enough to detect Floquet chiral edge modes and degenerate zero modes. In Fig.~$9$ we show the numerical results for the edge states in two cases with 3 bands and 4 bands, using $9$ or $12$ lattice sites. It is seen that in a relatively short lattice the edge states have more than $90\%$ of their probability distributed on the first three or four lattice sites at the boundaries. In particular, for the 3-band case, the number of edge states within each gap for a 9-site lattice under OBC is found to be the same as that for a very long lattice under OBC, which is also consistent with the bulk band Chern number.  In the 4-band case, almost degenerate modes close to zero quasi-energy are also found despite the fact that we considered only 12 lattice sites.
 Hence, in experiments, a short lattice already suffices.  Edge modes, once excited, might be imaged by suddenly switching off the main harmonic confinement and letting the atoms expand along the lattice, then detecting the spatial distribution of the atoms using absorption imaging~\cite{ExpSetRoati2008}.
Possible Floquet topological phase transitions might be detected via the concept of quantized adiabatic pumping in periodically driven systems~\cite{FQHDerekPRL2012}.  This can be a demanding task because in driven systems, one cannot count on the existence of a Fermi energy to uniformly fill a band.    Currently,  we are considering the pumping of some easier-to-prepare initial states in order to better detect the topological phase transitions in a periodically driven system.

\section{Concluding Remarks}

In this work it has been shown that a variety of topological states of matter may be generated using the CDHM, a rather simple $1$D system. Under the PBC, the eigenphase spectrum of the system forms Floquet bands with nonzero Chern numbers. Topological phase transitions are numerically demonstrated by changing the hopping amplitude, the superlattice potential and the driving period of the system. A Wannier state prepared on a single band which is well separated from the other bands is shown to move through a quantized number of lattice sites within an adiabatic cycle. In $3$-band and $5$-band cases, very flat bands are shown to emerge around eigenphase zero under certain choices of system parameters. In the $4$-band case, two Floquet bands are found to intersect conically at zero eigenphase, forming Dirac cones. Wavepackets prepared on these cones are shown to move non-dispersively in the lattice.

Under the OBC, we numerically observed that the nontrivial topology of Floquet bands manifests itself as chiral edge modes traversing the bulk gap. In $3$-band and $5$-band cases, anomalous edge modes with opposite chiralities are found to emerge on the same edge of the system.  In the $4$-band case, topologically protected degenerate edge modes appear at eigenphase zero.

In future work, it will be interesting to use the flat band as a starting point to study the interplay between topology and interaction in Floquet systems by adding self-interaction terms to the model.  It will also be interesting to consider if the degenerate zero modes may be utilized for quantum information applications. Lastly, since the model was also studied in the context of quantum chaos, it should be interesting to further examine the possible connection between the regular-to-chaos transition in the underlying classical limit and the topological phase transitions in the quantum system.
\section{Acknowledgments}
Long-Wen Zhou thanks Qifang Zhao and Cheng Shen for helpful discussions.
%

\appendix
\section{The Chiral symmetry of CDHM}

The Floquet operator for CDHM at a fixed phase shift $\beta$ is defined as:
\begin{equation}
\hat{U}_{\beta}(T,0) = {\bf T}e^{-i \int^T_0 dt \left[ J \text{cos}(\hat{k}) + V \text{cos}(2 \pi \alpha \hat{m} - \beta) \text{cos}(\Omega t) \right]}
\end{equation}
where we choose $\alpha = p/q$ with $p, q$ being co-prime integers. The driving frequency $\Omega = 2\pi/T$. In the main text we observed that the spectrum of $\hat{U}_{\beta}(T,0)$ with respect to $\beta$ has mirror symmetry with respect to the eigenphase zero and translational symmetry over $\pi$ in the $\beta$ direction. These properties can be understood by analyzing the chiral symmetry of CDHM.

In the main text, we define the chiral symmetry operator as $\Gamma_{\text{CDHM}} = e^{i\hat{m}\pi}$. This operator is unitary and square to be $1$, which means that:
\begin{equation}
\Gamma_{\text{CDHM}} = \Gamma^{-1}_{\text{CDHM}} = \Gamma^{\dagger}_{\text{CDHM}}
\end{equation}
Here $\hat{m}$ is the position operator which can only take integer values. This operator then performs a local unitary transformation in the position space. In the following we will prove the existence of two characteristics of the CDHM Floquet operator:
\begin{equation}
\Gamma_{\text{CDHM}} \hat{U}_{\beta}(T,0) \Gamma_{\text{CDHM}} = \hat{U}^{\dagger}_{\beta+\pi}(T,0)
\label{eqi}
\end{equation} and
\begin{equation}
\Gamma_{\text{CDHM}} \hat{U}_{\beta}(3T/4,-T/4) \Gamma_{\text{CDHM}} = \hat{U}^{\dagger}_{\beta}(3T/4,-T/4).
\label{eqii}
\end{equation}

To prove Eq.~(\ref{eqi}), we start with its left hand side, which can be expanded as:
\begin{equation}
\begin{split}
& \Gamma_{\text{CDHM}} \hat{U}_{\beta}(T,0) \Gamma_{\text{CDHM}} = \left[ e^{i\frac{J{\Delta}t}{2}\text{cos}(\hat{k})} e^{-iV{\Delta}t\text{cos}(2\pi\alpha\hat{m}-\beta)\text{cos}\left[\Omega(N-\frac{1}{2}){\Delta}t\right]} e^{i\frac{J{\Delta}t}{2}\text{cos}(\hat{k})} \right]\cdot \qquad\qquad \\
& \qquad\qquad\qquad\qquad\qquad\,\,\,\,\left[ e^{i\frac{J{\Delta}t}{2}\text{cos}(\hat{k})} e^{-iV{\Delta}t\text{cos}(2\pi\alpha\hat{m}-\beta)\text{cos}\left[\Omega(N-\frac{3}{2}){\Delta}t\right]} e^{i\frac{J{\Delta}t}{2}\text{cos}(\hat{k})} \right]\cdot... \\
& \qquad\qquad\qquad\qquad\qquad\,\cdot\left[ e^{i\frac{J{\Delta}t}{2}\text{cos}(\hat{k})} e^{-iV{\Delta}t\text{cos}(2\pi\alpha\hat{m}-\beta)\text{cos}(\Omega\frac{3}{2}{\Delta}t)} e^{i\frac{J{\Delta}t}{2}\text{cos}(\hat{k})} \right] \\
& \qquad\qquad\qquad\qquad\qquad\,\cdot\left[ e^{i\frac{J{\Delta}t}{2}\text{cos}(\hat{k})} e^{-iV{\Delta}t\text{cos}(2\pi\alpha\hat{m}-\beta)\text{cos}(\Omega\frac{1}{2}{\Delta}t)} e^{i\frac{J{\Delta}t}{2}\text{cos}(\hat{k})} \right] \\
\end{split}
\end{equation}
In the product we let $N\rightarrow\infty$, ${\Delta}t\rightarrow0$ and keep $N{\Delta}t = T$. On the right hand side we similarly have:
\begin{equation}
\begin{split}
& \hat{U}^{\dagger}_{\beta+\pi}(T,0) = \left[ e^{i\frac{J{\Delta}t}{2}\text{cos}(\hat{k})} e^{-iV{\Delta}t\text{cos}(2\pi\alpha\hat{m}-\beta)\text{cos}(\Omega\frac{1}{2}{\Delta}t)} e^{i\frac{J{\Delta}t}{2}\text{cos}(\hat{k})} \right]\cdot \qquad\qquad\qquad\qquad\qquad \\
& \qquad\qquad\quad\,\,\,\,\left[ e^{i\frac{J{\Delta}t}{2}\text{cos}(\hat{k})} e^{-iV{\Delta}t\text{cos}(2\pi\alpha\hat{m}-\beta)\text{cos}(\Omega\frac{3}{2}{\Delta}t)} e^{i\frac{J{\Delta}t}{2}\text{cos}(\hat{k})} \right]\cdot... \\
& \qquad\qquad\quad\,\cdot\left[ e^{i\frac{J{\Delta}t}{2}\text{cos}(\hat{k})} e^{-iV{\Delta}t\text{cos}(2\pi\alpha\hat{m}-\beta)\text{cos}\left[\Omega(N-\frac{3}{2}){\Delta}t\right]} e^{i\frac{J{\Delta}t}{2}\text{cos}(\hat{k})} \right] \\
& \qquad\qquad\quad\,\cdot\left[ e^{i\frac{J{\Delta}t}{2}\text{cos}(\hat{k})} e^{-iV{\Delta}t\text{cos}(2\pi\alpha\hat{m}-\beta)\text{cos}\left[\Omega(N-\frac{1}{2}){\Delta}t\right]} e^{i\frac{J{\Delta}t}{2}\text{cos}(\hat{k})} \right] \\
\end{split}
\end{equation}
Now it is easy to see that at each position of the product the left hand side equals the right hand side, since for any $M$ belongs to $[0,N]$ we have $\cos\left[\Omega (N-M) \Delta t\right] = \cos(\Omega M \Delta t)$. It is not hard to show that the relation (\ref{eqi}) is also true under the OBC.
 For the Floquet spectrum, this simply means that given an eigenstate $\ket{\psi}$ at any $\beta$ with eigenphase $\omega$, there must be another eigenstate $\Gamma_{\text{CDHM}}\ket{\psi}$ at $\beta+\pi$ with eigenphase $-\omega$.

To prove Eq.~(\ref{eqii}), we first expand its left hand side as:
\begin{equation}
\begin{split}
& \Gamma_{\text{CDHM}} \hat{U}_{\beta}(3T/4,-T/4) \Gamma_{\text{CDHM}} = \left[ e^{i\frac{J{\Delta}t}{2}\text{cos}(\hat{k})} e^{-iV{\Delta}t\text{cos}(2\pi\alpha\hat{m}-\beta)\text{cos}\left[\Omega(\frac{3N}{4}-\frac{1}{2}){\Delta}t\right]} e^{i\frac{J{\Delta}t}{2}\text{cos}(\hat{k})} \right]\cdot \\
& \qquad\qquad\qquad\qquad\qquad\qquad\qquad\,\left[ e^{i\frac{J{\Delta}t}{2}\text{cos}(\hat{k})} e^{-iV{\Delta}t\text{cos}(2\pi\alpha\hat{m}-\beta)\text{cos}\left[\Omega(\frac{3N}{4}-\frac{3}{2}){\Delta}t\right]} e^{i\frac{J{\Delta}t}{2}\text{cos}(\hat{k})} \right]\cdot... \\
& \qquad\qquad\qquad\qquad\qquad\qquad\quad\,\,\,\,\cdot\left[ e^{i\frac{J{\Delta}t}{2}\text{cos}(\hat{k})} e^{-iV{\Delta}t\text{cos}(2\pi\alpha\hat{m}-\beta)\text{cos}\left[\Omega(-\frac{N}{4}+\frac{3}{2}){\Delta}t\right]} e^{i\frac{J{\Delta}t}{2}\text{cos}(\hat{k})} \right] \\
& \qquad\qquad\qquad\qquad\qquad\qquad\quad\,\,\,\,\cdot\left[ e^{i\frac{J{\Delta}t}{2}\text{cos}(\hat{k})} e^{-iV{\Delta}t\text{cos}(2\pi\alpha\hat{m}-\beta)\text{cos}\left[\Omega(-\frac{N}{4}+\frac{1}{2}){\Delta}t\right]} e^{i\frac{J{\Delta}t}{2}\text{cos}(\hat{k})} \right] \\
\end{split}
\end{equation}
On the right hand side we have:
\begin{equation}
\begin{split}
& \hat{U}^{\dagger}_{\beta}(3T/4,-T/4) = \left[ e^{i\frac{J{\Delta}t}{2}\text{cos}(\hat{k})} e^{iV{\Delta}t\text{cos}(2\pi\alpha\hat{m}-\beta)\text{cos}\left[\Omega(\frac{N}{4}-\frac{1}{2}){\Delta}t\right]} e^{i\frac{J{\Delta}t}{2}\text{cos}(\hat{k})} \right]\cdot \qquad\qquad\qquad \\
& \qquad\qquad\qquad\qquad\,\,\left[ e^{i\frac{J{\Delta}t}{2}\text{cos}(\hat{k})} e^{iV{\Delta}t\text{cos}(2\pi\alpha\hat{m}-\beta)\text{cos}\left[\Omega(\frac{N}{4}-\frac{3}{2}){\Delta}t\right]} e^{i\frac{J{\Delta}t}{2}\text{cos}(\hat{k})} \right]\cdot... \\
& \qquad\qquad\qquad\quad\,\,\,\,\,\cdot\left[ e^{i\frac{J{\Delta}t}{2}\text{cos}(\hat{k})} e^{iV{\Delta}t\text{cos}(2\pi\alpha\hat{m}-\beta)\text{cos}\left[\Omega(\frac{N}{4}+\frac{3}{2}){\Delta}t\right]} e^{i\frac{J{\Delta}t}{2}\text{cos}(\hat{k})} \right] \\
& \qquad\qquad\qquad\quad\,\,\,\,\,\cdot\left[ e^{i\frac{J{\Delta}t}{2}\text{cos}(\hat{k})} e^{iV{\Delta}t\text{cos}(2\pi\alpha\hat{m}-\beta)\text{cos}\left[\Omega(\frac{N}{4}+\frac{1}{2}){\Delta}t\right]} e^{i\frac{J{\Delta}t}{2}\text{cos}(\hat{k})} \right] \\
\end{split}
\end{equation}
The left hand side and right hand side then equal with each other since for every term in the same position of the product and any $M$ belongs to $[0,N]$, we have $\cos\left[\Omega (3N/4-M) \Delta t\right] = -\cos\left[\Omega (N/4-M) \Delta t\right]$. Again, it can be shown that relation (\ref{eqii}) also holds under OBC. This result means that given an eigenstate $\ket{\psi}$ at any $\beta$ with the eigenphase $\omega$, then there must be another eigenstate $\Gamma_{\text{CDHM}}\ket{\psi}$ at the same $\beta$ with the eigenphase $-\omega$.

Relations ~(\ref{eqi}) and (\ref{eqii}) show the chiral symmetry of CDHM under two different choices of the symmetric time frame~\cite{BBCAsbothPRB2013}.   Finally, noting that a different choice of the starting time to express a Floquet operator cannot change its spectrum, it becomes obvious that $\hat{U}_{\beta}(T,0)$ and $\hat{U}_{\beta}(3T/4,-T/4)$ share the same eigenphase spectrum.  One can then conclude that in the eigenphase spectrum of CDHM with respect to the phase shift $\beta$,  it has a reflection symmetry with respect to eigenphase $0$ and $\pi$-translational symmetry along $\beta$-direction. Our numerical observations have confirmed these spectral features.

\section{The origin of the even Chern number of all bands for odd $q$}

From the Chern number table in the main text, we see that in $3$-band and $5$-band cases, all bands have even Chern numbers. To understand this, we rewrite our model Hamiltonian in the following form:
\begin{equation}
\begin{split}
& \hat{H}(t) = \sum_{m} \left[ \frac {J}{2} \left( \ket{m}\bra{m+1} + h.c. \right) + \frac {V}{2} \ket{m}\bra{m}\cos(2 \pi \alpha m - \beta + \Omega t) \right]\\
& \qquad\,+ \sum_{m} \left[ \frac {V}{2} \ket{m}\bra{m}\cos(2 \pi \alpha m - \beta - \Omega t) \right]\\
\end{split}
\end{equation}
Shifting time by $\beta/\Omega$, we find:
\begin{equation}
\begin{split}
& \hat{H}(t+\beta/\Omega) = \sum_{m} \left[ \frac {J}{2} \left( \ket{m}\bra{m+1} + h.c. \right) + \frac {V}{2} \ket{m}\bra{m}\cos(2 \pi \alpha m + \Omega t) \right]\\
& \qquad\qquad\quad\,+ \sum_{m} \left[ \frac {V}{2} \ket{m}\bra{m}\cos(2 \pi \alpha m - 2\beta - \Omega t) \right]\\
\end{split}
\end{equation}
For the Floquet operator under this time shift, it is related to the old one by a unitary transformation:
\begin{equation}
\hat{U}_{\beta}(T+\beta/\Omega,\beta/\Omega) = \hat{U}_{\beta}(\beta/\Omega,0) \hat{U}_{\text{CDHM}} \hat{U}_{\beta}(0,\beta/\Omega)
\end{equation}
This transformation only depends on $\beta/\Omega$ explicitly and thus will not affect the Chern number of the bands. To show this, denote $\hat{U}_{\beta}(\beta/\Omega,0)$ as $\hat{U}(\beta)$ and the eigenstate of $\hat{U}_{\text{CDHM}}(T,0)$ on band $n$ at a BZ point $(\phi,\beta)$ as $\ket{\psi_{n}(\phi,\beta)}$. Using an explicit Berry curvature expression for the band Chern numbers \cite{FQHDerekPRL2012},  the Chern number $C'_n$ of the $n$th Floquet band for the new operator $\hat{U}_{\beta}(T+\beta/\Omega,\beta/\Omega)$ is related to the $n$th's band Chern number $C_n$ of the propagator $\hat{U}_{\text{CDHM}}$ as:
\begin{equation}
\begin{split}
& C'_n = \frac {i}{2\pi} \int^{2\pi}_{0}\,d\beta \int^{2\pi}_{0}\,d\phi\,\,\{\langle \partial_{\phi} [\hat{U}(\beta)\psi_{n}(\phi,\beta)] | \partial_{\beta} [\hat{U}(\beta)\psi_{n}(\phi,\beta)] \rangle - h.c.\} \qquad\qquad\quad\\
& \quad\,= C_n + \frac {i}{2\pi} \int^{2\pi}_{0}\,d\beta \int^{2\pi}_{0}\,d\phi\,\,[\langle \partial_{\phi} \psi_{n}(\phi,\beta) | \hat{U}^{\dagger}(\beta)\partial_{\beta}\hat{U}(\beta) | \psi_{n}(\phi,\beta) \rangle - h.c.] \\
& \quad\,= C_n + \frac {i}{2\pi} \int^{2\pi}_{0}\,d\beta \int^{2\pi}_{0}\,d\phi\,\,\partial_{\phi} \langle \psi_{n}(\phi,\beta) | \hat{U}^{\dagger}(\beta)\partial_{\beta}\hat{U}(\beta) | \psi_{n}(\phi,\beta) \rangle \\
& \quad\,= C_n + \frac {i}{2\pi} \int^{2\pi}_{0}\,d\beta\,\,\langle \psi_{n}(\phi,\beta) | \hat{U}^{\dagger}(\beta)\partial_{\beta}\hat{U}(\beta) | \psi_{n}(\phi,\beta) \rangle|^{\phi=2\pi}_{\phi=0} \\
\end{split}
\end{equation}
The last term on the right hand side equals zero due to the single valuedness of the eigenstate at the same BZ point. This indicates that for each Floquet band, the operator $\hat{U}_{\beta}(T+\beta/\Omega,\beta/\Omega)$ and $\hat{U}_{\text{CDHM}}$ have exactly the same Chern numbers.

From the new Floquet operator we can see that if we scan $\beta$ from $0$ to $2\pi$, $2\beta$ will goes from $0$ to $4\pi$. So in this representation, the Chern number calculation previously defined can be regarded as a Berry curvature integral over a BZ of size $[0,2\pi)\times[0,4\pi)$, which is just two copies of the standard one. Because in the standard BZ of size $[0,2\pi)\times[0,2\pi)$, a Floquet band must have an integer Chern number, then this new representation exposing two copies of the standard BZ indicate that the same band must have a Chern number equal to an even integer.

\end{document}